\documentclass[twocolumn]{article}
\usepackage[utf8]{inputenc}
\usepackage[dvipdfmx]{graphicx}
\usepackage{siunitx}
\usepackage{authblk}
\usepackage{hyperref}
\usepackage{amsmath}

\let\Im\relax
\DeclareMathOperator{\Im}{Im}
\usepackage{booktabs}
\usepackage{graphicx}

\hypersetup{
    colorlinks=true,
    linkcolor=blue,
    filecolor=magenta,      
    urlcolor=cyan,
    }

\usepackage[margin=2.5cm]{geometry}

\title{\vspace{-15mm} High-resolution pressure imaging via background-oriented schlieren tomography: a spatiotemporal measurement for MHz ultrasound fields and hydrophone calibration}

\author[1]{\small{Sayaka Ichihara}}
\author[2]{Masato Yamagishi}
\author[1]{Yuta Kurashina}
\author[2]{Masanori Ota}
\author[1]{Yoshiyuki Tagawa}

\affil[1]{Department of Mechanical Systems Engineering, Tokyo University of Agriculture and Technology, Koganei Campus 6-204, 2-24-16 Nakacho, Koganei, Tokyo, Japan
}

\affil[2]{Department of Engineering, Chiba University, 1-33,Yayoi-cho, Inage-ku, Chiba, Japan
}

\date{}

\begin{document}

\twocolumn[

\maketitle

\paragraph{Abstract} 

In this work, the spatiotemporal pressure field of MHz-focused ultrasound is measured using a background-oriented schlieren technique combined with fast checkerboard demodulation and vector tomography (VT-BOS). Hydrophones have been commonly employed to directly measure the local pressure in underwater ultrasound. However, their limitations include that they disturb the acoustic field and affect the measured pressure through the spatial averaging effect. To overcome such limitations, we propose VT-BOS as a non-contact technique for acoustic field measurements using only a background image and a camera. In our experiments, VT-BOS measures focused acoustic fields with a focal width of 1.0 mm and a frequency of 4.55 MHz, capturing traveling, reflected, and standing waves. We discuss three key features of this approach: (1) the temporal evolution of pressure measured by VT-BOS and hydrophones, (2) the differences in computational cost and spatial resolution between VT-BOS and other techniques, and (3) the measurement range of VT-BOS. The results demonstrate that VT-BOS successfully quantifies spatiotemporal acoustic fields and can estimate the hydrophones' spatial averaging effect over a finite area. VT-BOS measures pressure fields of several MPa with high spatiotemporal resolution, requiring less computational and measurement time. It is used to measure pressure amplitudes from 0.4 to 6.4 MPa, with the potential to extend the range to 0.3–201.6 MPa by adjusting the background-to-target distance. VT-BOS is a promising tool for measuring acoustic pressure in the MHz and MPa ranges, critical for applications such as vessel flow measurement and hydrophone calibration.

\vspace{10pt}

\textbf{\textit{keyword--}}
Non-contact measurement technique, \ Acoustic Focused pressure field, \ MHz and MPa ultrasound, \ Background-oriented schlieren, \ Spatiotemporal measurement, \ Spatial averaging effects of hydrophone
%

\vspace{20pt}
]

\section{Introduction}\label{sec:intro}

\begin{figure}[ht]
	\centering
	\includegraphics[width=1.0\columnwidth]{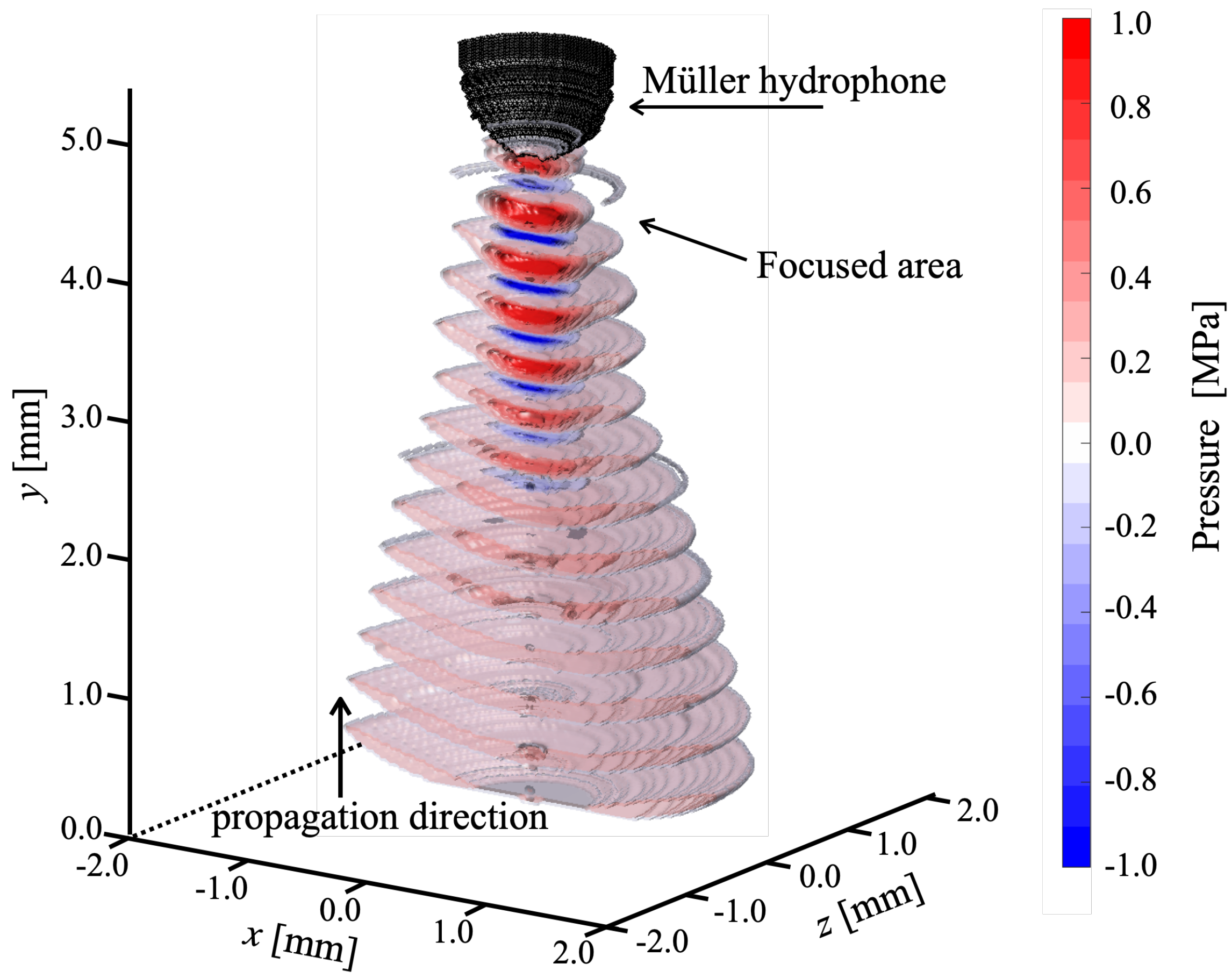}
	\caption{The three-dimensional ultrasound pressure field measured using the proposed VT-BOS technique.
    }
	\label{fig:vt_3D}
\end{figure}

Ultrasound has long been employed in various medical applications, including vessel flow measurement, cancer treatment, and and so on.
A vessel flow measurement technique uses focused ultrasound at frequencies in the MHz range and pressures of 0.2–4 MPa \cite{poindexter2011acoustic}.
For acoustic streaming, focused ultrasound operating at 1.5 MHz and pressures of 1–4 MPa can modify the temperature field and induce thermal lesions near blood vessels \cite{solovchuk2012simulation}.
In cancer treatment, focused ultrasound at frequencies of 1–2 MHz and amplitudes of 1–10 MPa facilitates the targeted delivery of drugs \cite{tachibana2001use,dayton2006application}.
Additionally, focused ultrasound can target a small focal region, on the order of millimeters, with temperatures exceeding 55°C, inducing the necrosis of cancer cells in a minimally invasive manner.

To effectively understand and control such ultrasound applications, it is crucial to measure the spatiotemporal acoustic field accurately.
Focused ultrasound operating at frequencies of several MHz and with maximum pressures of several MPa is key to many critical techniques \cite{ter2007high,giles2020magnetic}.


Hydrophones have commonly been used to measure the pressure fields of underwater ultrasound.
A hydrophone provides a direct local pressure measurement technique for measuring the time variation of pressure with high temporal resolution (e.g., 1 ns). 
However, hydrophones have three main issues.
First, they may disturb the pressure field during measurement \cite{saheban2021hydrophones}.
Second, when the hydrophone diameter exceeds the wavelength of the target ultrasound, the measured pressure is affected by the effect of spatial averaging over a finite local area in what is called the “spatial averaging effect" \cite{harris1985discussion,smith1989hydrophones,wear2020hydrophone}.
Third, measurements of the ultrasound pressure field by a hydrophone with high spatial resolution require a long measurement time since the hydrophone needs to scan the pressure field, i.e., it must be moved to different positions to measure the pressure to capture the overall pressure field \cite{kothapalli2017acoustic,nakamura2018quantitative}.

Image-based, non-contact measurement techniques offer the advantages of non-invasiveness and the ability to measure spatial distributions.
Techniques like optical phase contrast imaging \cite{nakamura2018quantitative, goldfain2021optical} and the schlieren technique  \cite{neumann2006schlieren, colom2023rapid, jiang2016quantitative} have gained attention as promising non-contact measurement approaches.
However, these techniques face challenges when measuring the spatiotemporal pressure field of focused ultrasound at several MPa and MHz frequencies.
Optical phase contrast imaging, for instance, is limited to measuring pressure amplitudes of up to 0.1 MPa due to optical diffraction effects.
The schlieren technique can be used to acquire the Laplacian density integrated along the optical axis in experiments, but the measurement error increases during the iterative calculation for determining the density field.

Here we focus on the background-oriented schlieren (BOS) technique \cite{meier2002computerized,venkatakrishnan2004density}.
BOS can measure pressure fields using only a background and a camera.
The technique involves capturing background images with and without the presence of the measurement target.
By comparing these two images, BOS visualizes the displacement field, which is proportional to the integrated density gradient along the camera axis.
Through three-dimensional reconstruction, BOS can determine the density gradient from the displacement field, which is then used to estimate the pressure field using an equation of state.
Recent advancements in displacement detection methods \cite{shimazaki2022background,cakir2023assessment,vinnichenko2023performance} and reconstruction techniques \cite{ichihara2022background,gao2023reconstruction,amjad2023three} have significantly enhanced the quantitative measurement capabilities of BOS.

In this study, we measure the acoustic pressure fields of underwater focused ultrasound at 4.554 MHz and several MPa with high spatiotemporal resolution. We utilize BOS enhanced by fast checkerboard demodulation (FCD), vector tomography (VT), and a high-resolution camera, as recently proposed by the authors \cite{shimazaki2022background, ichihara2022background}.
This proposed BOS method (called VT-BOS hereafter) can characterize the pressure field with a high spatial resolution of 1 µm on image and a high temporal resolution of 15 ns.
Figure \ref{fig:vt_3D} shows the three-dimensional acoustic pressure field measured using VT-BOS at a particular moment in our experiment.
First, we compare the VT-BOS measurement results with those measured by two hydrophones.
The spatial averaging effect of the hydrophones is discussed quantitatively. 
Second, we evaluate the performance of VT-BOS against hydrophones, conventional BOS, and other non-contact techniques based on key factors such as spatial resolution and computational cost.
Lastly, we estimate the measurable pressure range of the VT-BOS method so that this technique can be further used to measure ultrasound fields.

In the rest of this section, we briefly review how VT-BOS addresses the limitations of conventional BOS in spatiotemporal field measurements by introducing a novel displacement detection method and reconstruction technique. 
Conventional BOS \cite{venkatakrishnan2005density} typically employs a cross-correlation-based particle image velocimetry (PIV) algorithm \cite{van2014density,kahler2016main} for displacement detection and filtered back-projection (FBP) \cite{amjad2020assessment,sourgen2012reconstruction} for reconstruction.
In conventional BOS, the Laplacian density is obtained by reconstructing the differentiated displacement, and the density is computed through iterative calculations.
Experimental errors, increased by the differentiation of the displacement and the reduced spatial resolution caused by PIV, significantly affect the convergence of the iterative calculations.
Therefore, conventional BOS is unsuitable for measuring shock waves and ultrasound with large pressure fluctuations, which require spatial resolution.
To overcome these issues, various approaches have been explored, including the application of AI \cite{luo2020rapid}, physical model assumptions \cite{koponen2022nonlinear}, and hydrophone corrections \cite{pulkkinen2017ultrasound} to conventional BOS.
Our proposed VT- BOS method integrates novel techniques: fast checker demodulation (FCD) as a displacement detection method \cite{shimazaki2022background} and vector tomography (VT) as a reconstruction technique \cite{ichihara2022background}.
FCD achieves pixel-level resolution in displacement measurement by detecting phase modulation in a distorted periodic pattern, outperforming PIV, which relies on the resolution of interrogation subregions from images.
Additionally, FCD can measure larger displacements than PIV.
The VT technique enables a direct vector field reconstruction by utilizing an axisymmetric assumption and linear matrix integration, as opposed to the differentiation and iterative calculation required by FBP.
This approach significantly reduces computational costs and enhances accuracy compared to FBP \cite{ichihara2022background}.
VT-BOS has demonstrated superior capability in measuring spatiotemporal pressure fields, surpassing other current innovations in BOS \cite{ichihara2022background}.

\section{Measurement principle}\label{sec:principle}
In this section, Section \ref{sec:BOS} explains the principles of VT-BOS, while Section \ref{sec:hydrophone} outlines the characteristics of hydrophones.

\subsection{Background-oriented schlieren technique}\label{sec:BOS}

\begin{figure}[ht]
	\centering
	\includegraphics[width=1.0\columnwidth]{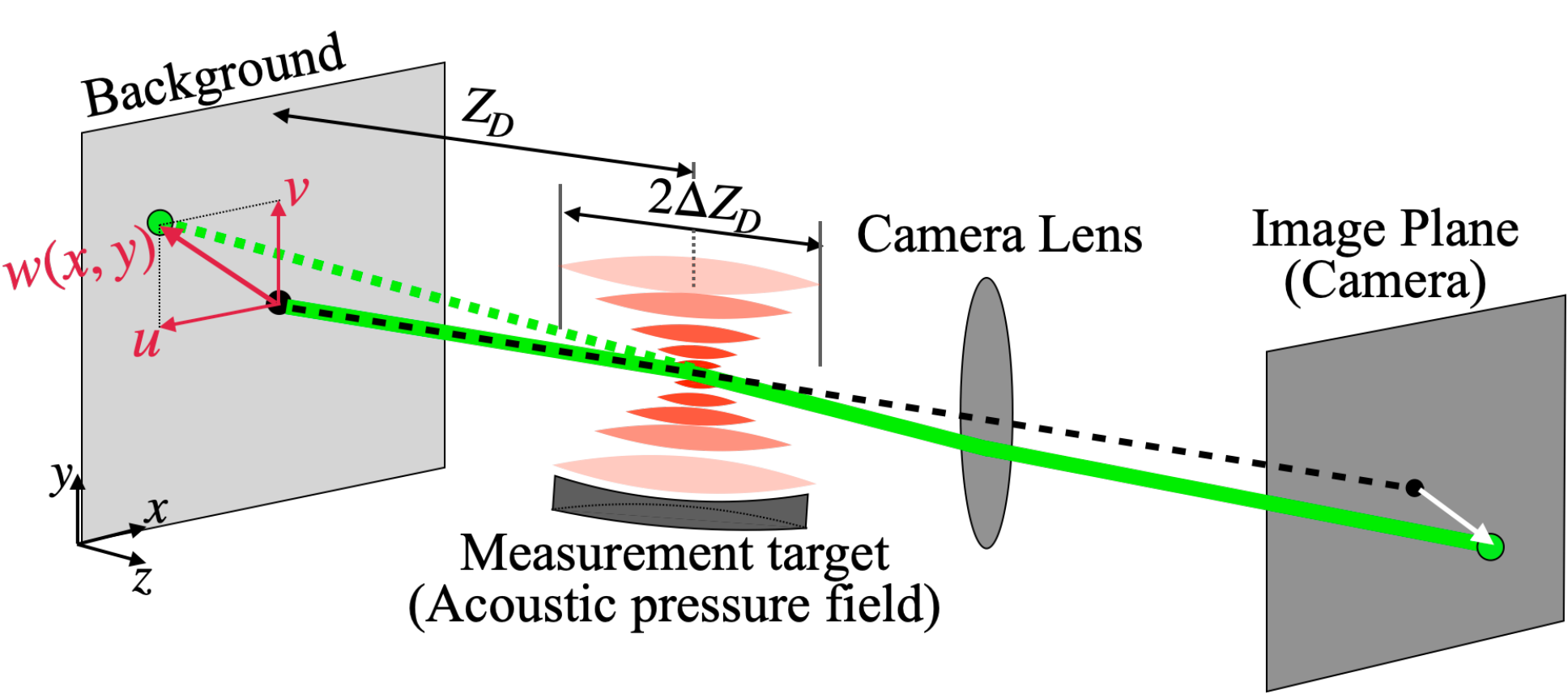}
	\caption{
    A schematic of the BOS arrangement.
    The camera used for capturing the background depicts the light rays as follows.
    A black dashed line represents a ray when the measurement target is absent. A green line shows a ray when the measurement target is included.
    A green dashed line indicates a ray estimated by a camera when the target is included.
    The displacement vector on the background is $\boldsymbol{w}(x,y) = (u,v)$.
    $Z_D$ is the distance from the center of the target to the background image and $2\Delta Z_D$ is the thickness of the target.
    }
	\label{fig:bos-theory}
\end{figure}

BOS uses only the background and a camera to measure the acoustic pressure field (the measurement target).
The optical setup is illustrated in Fig. \ref{fig:bos-theory}.
The camera is focused on the background.
When a background image is captured without a target (in what is referred to as the reference image), the black dot on the center of the background is projected onto the image plane along the dashed black line.
However, when the target is present, the light ray from a point on the background (the black dot in Fig. \ref{fig:bos-theory}) is refracted due to the difference in refractive index between the target and the surrounding fluid.
If the target is positioned along the dashed black line between the camera and the background, as shown in Fig. \ref{fig:bos-theory}, the point on the background (the black dot) is projected onto the image plane, passing through the refracted ray (the rigid green line).
Although the pattern on the background remains unchanged, the pattern projected onto the image plane captured by the camera looks distorted by the presence of the target (and is referred to as the distorted image). This apparent distortion, as shown by the red arrows in Fig. 2, represents the displacement $u$ in the $x-$direction and the displacement $v$ in the $y-$direction. The displacement vector on the background $\boldsymbol{w}$, where $\boldsymbol{w} = (u, v)$, is obtained by comparing the reference and distorted images using the displacement detection method, which is detailed in Section \ref{sec:fcd}.

The displacement vector $\boldsymbol{w}$ contains information about the spatial gradient of the refractive index integrated along the $z$-axis \cite{venkatakrishnan2005density}, expressed as:
\begin{align}
    u = \frac{Z_D}{n_0}\int^{Z_D + \Delta Z_D}_{Z_D - \Delta Z_D}\frac{\partial n}{\partial x} dz 
    = \frac{K Z_D}{n_0} \int \frac{\partial \rho}{\partial x} dz, 
    \label{eq:bos_u} 
\end{align}
\begin{align}
    v = \frac{Z_D}{n_0}\int^{Z_D + \Delta Z_D}_{Z_D - \Delta Z_D}\frac{\partial n}{\partial y} dz 
    = \frac{K Z_D}{n_0} \int \frac{\partial \rho}{\partial y} dz, 
    \label{eq:bos_v}
\end{align}
where $n$ is the refractive index of the target, $n_0$ is the refractive index of the surrounding fluid, $Z_D$ is the distance from the center of the target to the background image, $2\Delta Z_D$ is the thickness of the target, and $K (= 3.14 \times 10^{-4} \ \rm{m^3 / kg})$ is the Gladstone–Dale constant \cite{ichihara2022background, van2014density} in the Gladstone–Dale equation:
\begin{gather}
    n = \rho K + 1.
    \label{eq:bos_gland-stone}
\end{gather}

Since the displacement field is related to the integrated density gradient field, reconstruction of the three-dimensional density gradient field is necessary. 
Section \ref{sec:vt} details the reconstruction technique used to derive the three-dimensional density gradient field and then the density field.
Once the density field is obtained, the pressure field can be calculated from the equation of state (Tait equation):

\begin{align}
    \frac{p + B}{p_0 + B} = \left( \frac{\rho }{\rho_0} \right)^{\alpha},
    \label{eq:bos_tait}
\end{align}
where $p_0 \ (= 101.3 \ \rm{kPa})$ is the atmospheric pressure, $\rho_0 (= 998 \ \rm{kg/m^3})$ is the density of the ambient fluid, and $B \ (= 314 \ \rm{MPa})$ and $\alpha \ (= 7)$ are constants in the case of the density of standard-state water \cite{brujan2010cavitation}.

\subsubsection{Displacement detection: fast checker demodulation}\label{sec:fcd}

\begin{figure*}[ht]
	\centering
	\includegraphics[width=1.0\textwidth]{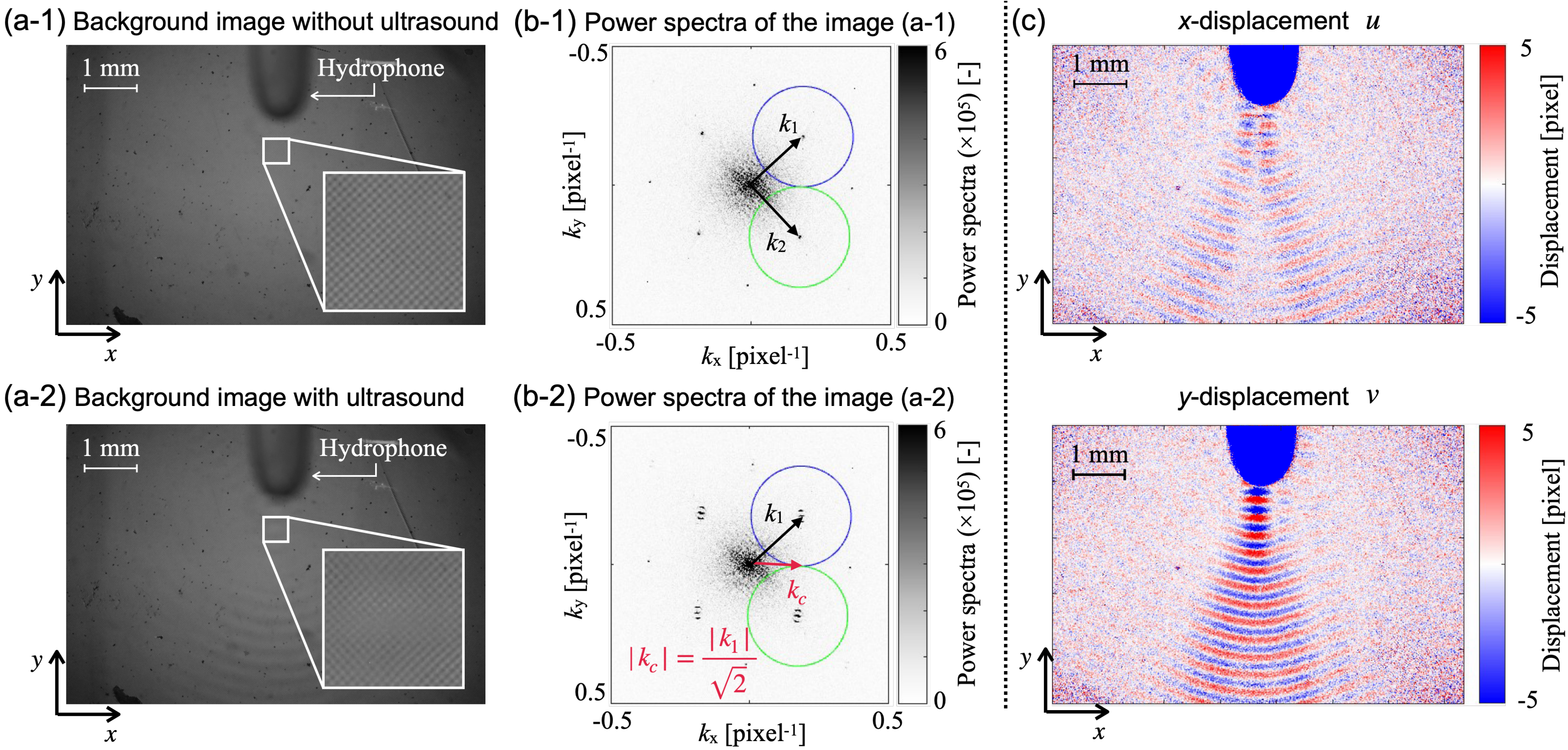}
	\caption{
 The BOS procedure for calculating the displacement field using fast checker demodulation (FCD). (a) The top image shows the checker background image without focused ultrasound. The bottom image shows the checker background image with focused ultrasound. (b) The power spectra in Fourier space ($k$-space) for background images without and with focused ultrasound are shown in the top and bottom figures, respectively. (c) The top figure represents the displacement field $u$, which is the magnitude of the $x$ component of the displacement vector $\boldsymbol{w}$, while the bottom figure represents the displacement field $v$. Both were detected using FCD.}
	\label{fig:fcd}
\end{figure*}

FCD is adopted as the displacement detection method in this work.
FCD is less susceptible to detection errors from missing patterns and can achieve pixel-level resolution by detecting phase modulation \cite{wildeman2018real}.
In the following section, we briefly review the principle of FCD as implemented by Wildman $\it{et \ al.}$ \cite{wildeman2018real} and Shimazaki $\it{et \ al.}$ \cite{shimazaki2022background}.
Note that BOS has traditionally used PIV, which relies on a random-dot pattern background, as a displacement detection method.
PIV detects displacements by matching the pattern in subdivided interrogation areas (typically 64–1024 pixels) of the reference image to the distorted images \cite{kahler2016main}.
However, PIV encounters two major issues. First, the spatial resolution of the displacement field is reduced compared to the reference image by an amount depending on the size of the interrogation area \cite{raffel2018particle}.
Second, detecting displacements in distorted images becomes impossible when dots are deformed or disappear due to large density gradients \cite{mckenna2002performance}.

The FCD calculation steps are schematically described in Fig. \ref{fig:fcd}.
FCD typically employs a periodic background pattern, such as a checkerboard (see Fig. \ref{fig:fcd} (a-1) and (a-2)) \cite{shimazaki2022background}.
To calculate displacement, FCD first obtains the Fourier space ($k$-space) of both the reference and distorted images using a fast Fourier transform (FFT), as shown in Fig. \ref{fig:fcd} (b-1) and (b-2).
The $k$-space contains several periodic power spectra components known as carrier peaks, as shown by the black area in Fig. \ref{fig:fcd} (b-1).
The carrier peaks in the distorted image, which arise from distortion, are separated from the carrier peaks in the reference image via Fourier domain filtering.
This separated carrier signal is related to the displacement (see Fig. \ref{fig:fcd} (c)).

The brightness distribution of the reference image $I_0(\boldsymbol{r})$ is represented as a complex form of Fourier series:
\begin{equation}
    I_0(\boldsymbol{r}) = \sum\limits_{m=-\infty}^{\infty} \sum\limits_{n=-\infty}^{\infty} a(m,n)e^{i(m\boldsymbol{k_1}+n\boldsymbol{k_2})\cdot \boldsymbol{r}},
    \label{eq:fcd-I0}
\end{equation}
where $\boldsymbol{k_1}$ and $\boldsymbol{k_2}$ are reciprocal lattice vectors, vector $\boldsymbol{r}$ indicates position in Cartesian coordinates $(x,y)$, and the $n$ and $m$ of the expansion coefficients $a(m,n)$ are natural numbers larger than 1.
The lattice vectors $\boldsymbol{k_1}$ and $\boldsymbol{k_2}$ are defined by the carrier peaks in $k$-space.
The distortion in the distorted image, represented as a displacement vector field $\boldsymbol{w}(\boldsymbol{r})$, modulates the phase of the Fourier series of the reference image $I_0(\boldsymbol{r})$.
Herein, the brightness of the distorted image is denoted as
\begin{multline}
    I = I_0(\boldsymbol{r}-\boldsymbol{w}(\boldsymbol{r})) \\
        =\sum\limits_{m=-\infty}^{\infty} \sum\limits_{n=-\infty}^{\infty} a(m,n)e^{i(m\boldsymbol{k_1}+n\boldsymbol{k_2})\cdot 
        ( \boldsymbol{r - \boldsymbol{w}(\boldsymbol{r})})}.
    \label{eq:fcd-I}
\end{multline}
Here, the carrier peak in the power spectra of the distorted image $I$, which contains the distortion information, is defined as $\boldsymbol{k_c}$.
If the displacement $\boldsymbol{w}(\boldsymbol{r})$ is not too large or too small compared with the wavelengths of the periodic pattern in the reference image $I_0(\boldsymbol{r})$, the carrier peak $\boldsymbol{k_c}$ will be localized around the carrier peaks of the periodic pattern in $k$-space: $\boldsymbol{k_c} \in m\boldsymbol{k_1} + n\boldsymbol{k_2}$ \cite{grediac2016grid, takeda1982fourier}.
To separate the displacement $\boldsymbol{w}(\boldsymbol{r})$ from these carrier peaks, FCD applies FFT to Eqs. \ref{eq:fcd-I0} and \ref{eq:fcd-I}:
\begin{align}
    F\left[ I_0(\boldsymbol{r}) \right] = 
    \int_{-\infty}^{\infty} \int_{-\infty}^{\infty} f(\boldsymbol{r})e^{-i \boldsymbol{k_c}\cdot \boldsymbol{r}} d\boldsymbol{r},
    \label{eq:fcd-FFT-I0}\\
    F\left[ I(\boldsymbol{r}) \right] =
    \int_{-\infty}^{\infty} \int_{-\infty}^{\infty} f(\boldsymbol{r})e^{-i \boldsymbol{k_c}\cdot( \boldsymbol{r} - \boldsymbol{w}(\boldsymbol{r}))} d\boldsymbol{r},
    \label{eq:fcd-FFT-I}
\end{align}
where $\boldsymbol{k_c} \in \boldsymbol{k_1} + \boldsymbol{k_2}$ to focus on the frequency related to the displacements $\boldsymbol{w}(\boldsymbol{r})$.
Note that the vectors $\boldsymbol{k_1} + \boldsymbol{k_2}$ at $m$ = 1 and $n$ = 1 correspond to the carrier peak at the fundamental frequency of the square wave, which emerges when a Fourier transform is applied to the square wave (representing the periodic background).

The noise (e.g., fluctuations in light intensity) appears at significantly higher frequencies than the displacement in $k$-space.
Fourier domain filtering, a low-pass filter, removes this noise from  Eqs. \ref{eq:fcd-FFT-I0} and \ref{eq:fcd-FFT-I}.
After filtering, the inverse Fourier transform is applied to Eqs. \ref{eq:fcd-FFT-I0} and \ref{eq:fcd-FFT-I}:
\begin{align}
    g_0(\boldsymbol{r}) = a_ce^{i \boldsymbol{k_c} \cdot \boldsymbol{r}}, 
    \label{eq:fcd_g0} \\
   g(\boldsymbol{r}) = a_ce^{i \boldsymbol{k_c} \cdot( \boldsymbol{r} - \boldsymbol{w}(\boldsymbol{r}))},
    \label{eq:fcd_g}
\end{align}
where $a_c$ is complex number and fluctuation 
component created in experiment.
Therefore, the displacement $\boldsymbol{w}$ is derived from the complex number $g$ (Eq. \ref{eq:fcd_g}) by multiplying it by the conjugate of $g_0$ (Eq. \ref{eq:fcd_g0}) and then applying the natural logarithm to obtain the resulting complex value:
\begin{equation}
    \phi(\boldsymbol{r}) \equiv 
    \Im{\left( \ln (gg_0^*) \right)} 
    = -\boldsymbol{k_c}\cdot\boldsymbol{w(\boldsymbol{r})}.
    \label{eq:fcd_phi}
\end{equation}
Herein, this relationship between the complex signal and displacement is as described by Takeda $\it{et \ al}$ \cite{takeda1982fourier} and Grediac $\it{et \ al}$. \cite{grediac2016grid}.
When using a checkerboard background image, $\boldsymbol{k_1}$ and $\boldsymbol{k_2}$ are linearly independent reciprocal vectors, $|\boldsymbol{k_1}| = |\boldsymbol{k_2}|$ and $\boldsymbol{k_1} \perp \boldsymbol{k_2}$.
By applying the condition for $\boldsymbol{k_c}$ to Eq. \ref{eq:fcd_phi}, two equations related to $\boldsymbol{k_1}$ and $\boldsymbol{k_2}$,
\begin{align}
    \phi_1(\boldsymbol{r}) = -\boldsymbol{k_1}\cdot\boldsymbol{w(\boldsymbol{r})}, \\
    \phi_2(\boldsymbol{r}) = -\boldsymbol{k_2}\cdot\boldsymbol{w(\boldsymbol{r})},
\end{align}
 can be obtained for each pixel.

To enhance the displacement detection performance, the modulated carrier peaks must be well separated in $k$-space.
To satisfy this assumption, FCD has two existing criteria and one criterion for avoiding phase wrapping, given as
\begin{align}
    k_s \ <& \ \frac{k_{c}}{\sqrt{2}},
    \label{Eq:fcd-criteria1} \\
     k_s u^{\prime}_s \ <& \ \frac{1}{\sqrt{2}},
    \label{Eq:fcd-criteria2} \\
    k_c u_s\  <& \ \pi,
    \label{Eq:fcd-criteria3}
\end{align}
where $k_c ( =|\boldsymbol{k_c}|) = |\boldsymbol{k_1}|/\sqrt{2}$ in the checker pattern background, $k_s$ is the wavenumber generated by the displacement, $u_s$ is the maximum amplitude of displacement, and $u^{\prime}_s$ is the maximum displacement gradient.
The conditions described in Eqs. \ref{Eq:fcd-criteria1} and \ref{Eq:fcd-criteria2} are applied specifically for cases using a checkerboard pattern background.
In our case, $k_s$ is the wavenumber of the measurement target, i.e., ultrasound, in the distorted image.
Equation  \ref{Eq:fcd-criteria1} defines the measurable wavelength of the target in relation to the wavelength of the background.
Equation \ref{Eq:fcd-criteria2} defines the limit of the maximum measurable displacement gradient in terms of the wavelength of the target.
When extracting displacement from the imaginary part of the complex logarithm (Eq. \ref{eq:fcd_phi}), the phase wraps at a period of $2\pi$, so if it exceeds $\pi, 3\pi, 5\pi$, etc., the phase inverts.
Equations 16 show the criteria for avoiding phase wrapping.
Note that Eqs. \ref{Eq:fcd-criteria1} and \ref{Eq:fcd-criteria2} were satisfied under the experimental conditions.

\subsubsection{Three-dimensional reconstruction of the density field: vector tomography}\label{sec:vt}

The displacement vector $\boldsymbol{w}$ is related to the integrated density gradient (Eqs. \ref{eq:bos_u} and \ref{eq:bos_v}).
The three-dimensional density gradient field is calculated using a reconstruction technique in BOS.
For axisymmetric measurement targets, BOS commonly utilizes filtered back projection (FBP) as a reconstruction technique \cite{sourgen2012reconstruction, amjad2020assessment, pulkkinen2017ultrasound}.
However, FBP presents several problems, such as strong dependence of the reconstruction accuracy on the spatial resolution, noise amplification during displacement differentiation, and high computational cost.
This work employs vector tomography (VT) \cite{ichihara2022background} to address these issues.
VT can directly reconstruct the three-dimensional distribution from the integrated values of the vector field using an inverse matrix (as shown later in Eq. \ref{Eq:VT-bos}).
As discussed in detail by Ichihara $\it{et} \ \it{al}$ \cite{ichihara2022background}, the calculation accuracy in VT is almost independent of the spatial resolution and is higher than that in FBP.
Furthermore, the computational cost of VT is less than 1/1800 of that required by FBP.

\begin{figure}[ht]
	\centering
	\includegraphics[width=1.0\columnwidth]{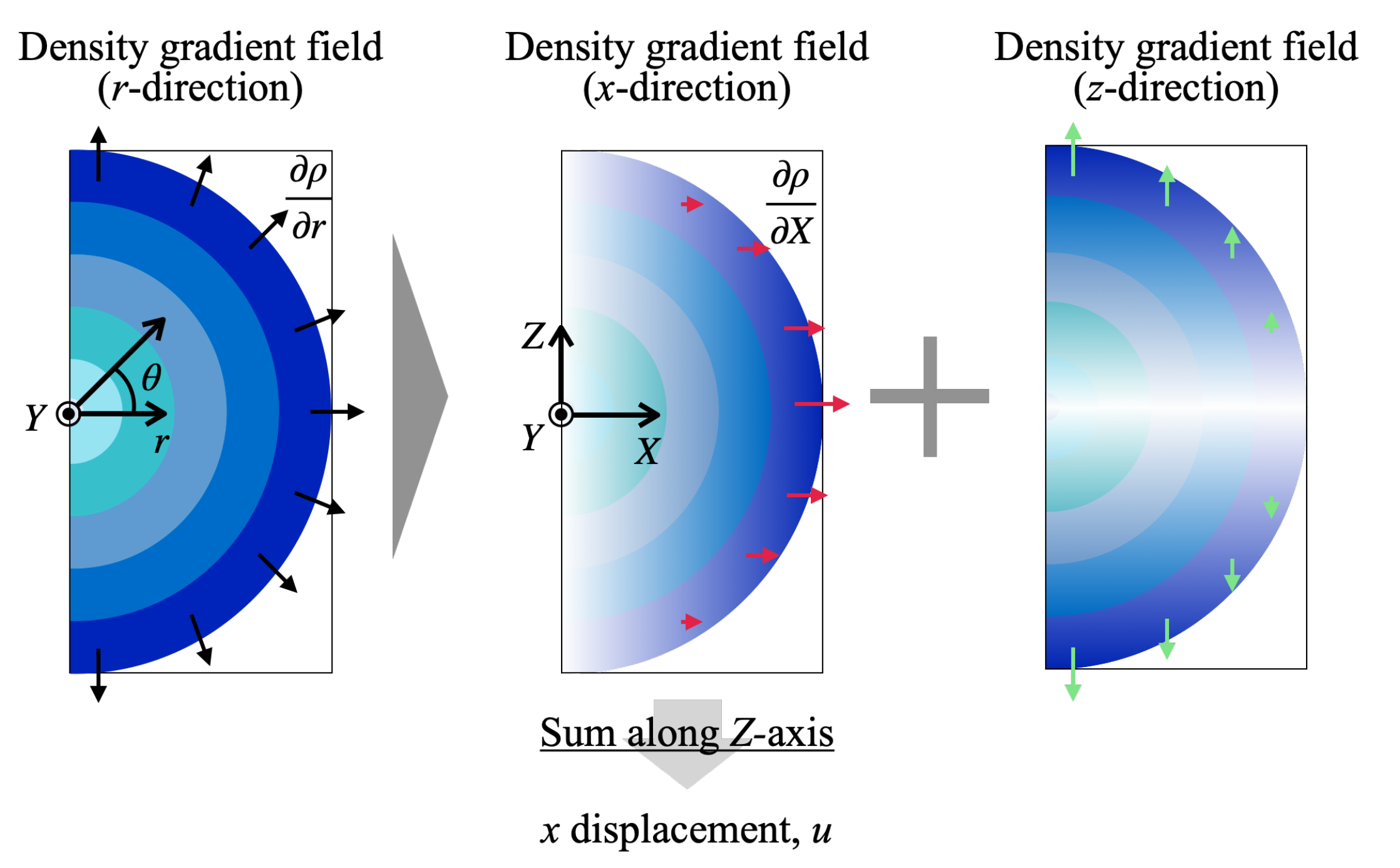}
	\caption{
    The relationship between the reconstruction field (density gradient field) and projection value (displacement) for an axisymmetric density gradient.
    The black vectors show the density gradient.
    The red and light blue vectors show the $x$ component and $z$ component of the density gradient, respectively.
    The color contour shows the magnitude of the vector in arbitrary units.
    }
	\label{fig:vt}
\end{figure}


To briefly review the VT method \cite{ichihara2022background}, it is important to determine the coordinate system used in the three-dimensional reconstruction field.
The reconstruction field employs Cartesian coordinates denoted by capital letters $(X, Y, Z)$.
The reconstructed distribution (measurement target) at the $X$-$Z$ cross-section is illustrated in Fig. \ref{fig:vt}.
The Cartesian coordinate system $(X, Y, Z)$ has its origin on the symmetric axis of the axisymmetric reconstructed distribution.
The cartesian coordinates $(x, y, z)$ used in the BOS setup (Fig. \ref{fig:bos-theory}) relate to $(X, Y, Z)$ as: $X = x$, $Y = y$,  and $Z = z + Z_D$ (where $Z_D$ is the constant distance described in Sec. \ref{sec:BOS}).

VT assumes that the density gradient field $\partial \rho / \partial r$ is axisymmetric around the $Y$-axis; therefore, $\partial \rho / \partial r$ does not vary in the circumferential direction ($\theta$-direction) but depends solely on the distance along the $r$-axis, as shown in Fig. \ref{fig:vt}.
The polar coordinates $(r, \theta)$ are related to the Cartesian coordinates $(X, Z)$ by the relations $X = r \cos \theta$ and $Z = r \sin \theta$.
As a result, the term $\partial \rho / \partial x$ in Eq. \ref{eq:bos_u} can be expressed as $\partial \rho / \partial r \cdot \cos \theta$:
\begin{multline}
    u(X) = \frac{KZ_D}{n_0}  \int^{\Delta Z_D}_{-\Delta Z_D} \frac{\partial}{\partial r} \rho \left( \sqrt{X^2 + Z^2} \right) \\ 
    \times \frac{X}{\sqrt{X^2 + Z^2}} dZ, 
    \label{Eq:VT1}
\end{multline}


To establish a connection between the displacement  $u(X)$ and the density gradient $\partial \rho(\sqrt{X^2 + Z^2}) / \partial r$ in matrix form, the integral equation (Eq. \ref{Eq:VT1}) is discretized.
The Cartesian coordinates $(X, Z)$ are replaced by discretized Cartesian coordinates $(\bar{X}, \bar{Y})$, where values exist only at equally spaced grid points.
Additionally, the integral interval $\Delta Z_D$ is replaced by a natural number $N$, which is the closest integer greater than or equal to $\Delta Z_D$, ensuring proper sampling resolution.

At $\bar{X}$ = 1,  the discretized version of Eq.\ref{Eq:VT1} in Cartesian coordinates is written as:
\begin{multline}
    u(1) = \frac{KZ_D}{n_0}\biggl\{
    \alpha_{1,1} \frac{\partial}{\partial r} \rho \left( \sqrt{2} \right) + \alpha_{1,2} \frac{\partial}{\partial r} \rho \left( \sqrt{5} \right) + \cdots \\
    +\alpha_{1,N} \frac{\partial}{\partial r} \rho \left( \sqrt{1+N^2} \right)
    \biggr\},
    \label{Eq:VT-example}
\end{multline}
where $\alpha_{\bar{X}, \bar{Z}} \propto \bar{X} / \sqrt{\bar{X}^2 + \bar{Z}^2} \left( = \cos \theta_{\bar{X}, \bar{Z}} \right)$.
For interpolation between real values (e.g. $\partial \rho ( \sqrt{2} ) / \partial r$) , linear interpolation is employed, using the two closest natural numbers to approximate the real values.
Therefore, the discretized $x$-displacement $u(\bar{X})$ in Eq. \ref{Eq:VT1} can be expressed as: 
\footnotesize
\begin{gather}
\begin{bmatrix}
u(1)\\
u(2)\\
u(3)\\
\vdots \\
u(N)\\
\end{bmatrix}
=  \frac{KZ_D}{n_0}
\begin{bmatrix}
\alpha_{1,1}&\alpha_{1,2}&\alpha_{1,3}&\cdots &\alpha_{1,N}\\
0&\alpha_{2,2}&\alpha_{2,3}&\cdots&\alpha_{2,N}\\
0&0&\alpha_{3,3}&\cdots&\alpha_{3,N}\\
\vdots&\vdots&\vdots&\ddots&\vdots\\
0&0&0&\cdots&\alpha_{N,N}
\end{bmatrix}
\begin{bmatrix}
\dfrac{\partial}{\partial r} \rho(1)\\
\dfrac{\partial}{\partial r} \rho(2)\\
\dfrac{\partial}{\partial r} \rho(3)\\
\vdots\\
\dfrac{\partial}{\partial r} \rho(N)
\end{bmatrix}
 \label{Eq:VT-bos}
\end{gather}
\normalsize
where the constant coefficient $\alpha_{\bar{X}, \bar{Z}}$ includes both the interpolation factor and  $\cos \theta_{\bar{X}, \bar{Z}}$.

The density gradient matrix $\partial \boldsymbol{\rho}/ \partial r$ at a certain $y=y_0$ is obtained by applying the inverse matrix of $\alpha_{i,j}$ to both sides of Eq. \ref{Eq:VT-bos}.  
Subsequently, the density gradient field in the $Y-r$ cross-section is obtained by repeating the aforementioned procedure for different $y$.
Finally, the density field can be calculated by integrating $\partial \rho/ \partial r$ along the $r$-axis,
\begin{align}
    \rho = \int^{\infty}_{0} \frac{\partial \rho}{ \partial r} dr,
    \label{Eq:bos_rhog}
\end{align}
  where the boundary condition in Eq. \ref{Eq:bos_rhog} is set as a water density of $\rho_0$ (= 998 kg/m$^3$) at $r = \infty$.

\subsection{Hydrophones}\label{sec:hydrophone}
\begin{figure}[ht]
	\centering
	\includegraphics[width=1.0\columnwidth]{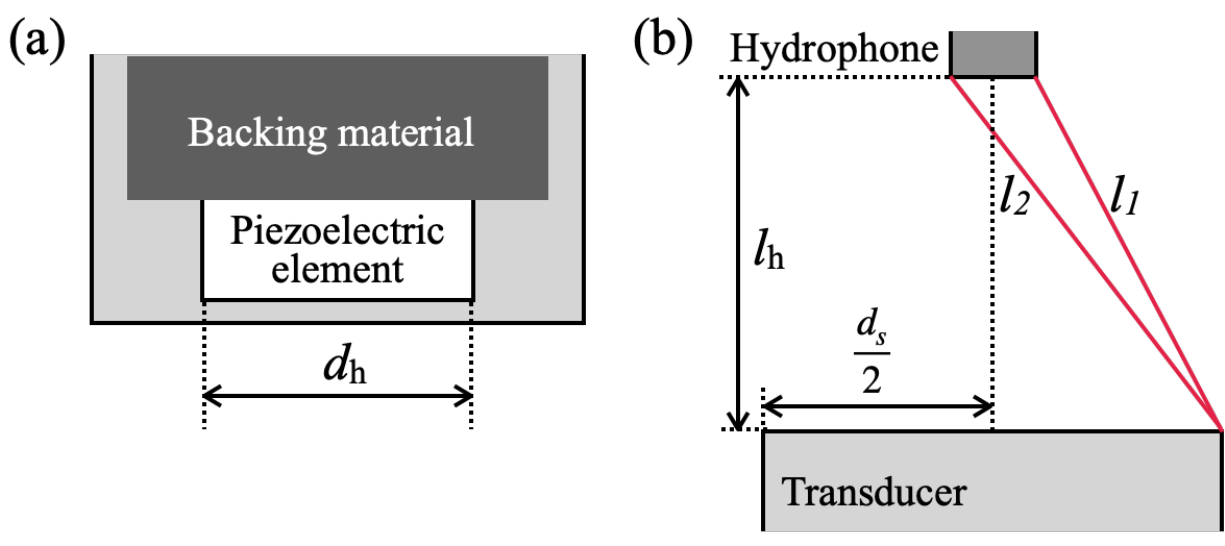}
	\caption{
A schematic of the hydrophone pressure measurement system.
    (a) An illustration of the internal structure of a hydrophone tip.
    (b) The spatial relationship between ultrasound emitted by a circular transducer and the measurement area of the hydrophone.
    }
	\label{fig:hydrophone}
\end{figure}

Hydrophones are among the most widely used underwater pressure measurement techniques.
One common form is the needle-type hydrophone, which utilizes a piezoelectric element as the pressure sensor.
A hydrophone measures local pressure by converting pressure fluctuations into electrical signals through a piezoelectric element.
The measured local pressure represents an average pressure over an area determined by the size of the hydrophone tip in the hydrophone.
When the tip size exceeds the characteristic length of the pressure wave, a “spatial averaging effect" occurs, which will be detailed in the following paragraph.
The internal structure of the hydrophone tip is illustrated in Fig. 5(a).
The piezoelectric element is housed within a metallic casing for protection.
A backing material is placed behind the piezoelectric element, allowing the hydrophone to measure across a broad frequency range rather than being limited to the resonant frequency.
Two piezoelectric materials are commonly used: piezoelectric ceramics and the piezoelectric polymer polyvinylidene fluoride (PVDF) \cite{xing2021review,zou2003wideband}.
In the experiment in this work, hydrophones with different piezoelectric elements were employed. Detailed information is provided in Section 3.1.

Hydrophones provide several advantages for measuring acoustic pressure, including a high-frequency response (in the tens of MHz), the ability to measure localized pressures based on sensor size (typically 0.3–1.0 mm), and the capacity to detect large pressure fluctuations ranging from MPa to GPa \cite{saheban2021hydrophones}.
Despite their widespread use in measuring acoustic pressure fields, hydrophones face challenges when dealing with high-intensity acoustic environments \cite{harris1985discussion,radulescu2001hydrophone}.
First, hydrophones require spatial scanning to map pressure fluctuations at various locations.
For instance, measuring a $4.0 \times 4.0$ mm$^2$ area with a spatial resolution of 0.1 mm necessitates 1600 data points, which could take approximately 3.5 hours to acquire.
Second, when measuring focused acoustic fields with pressures in the MPa range and frequencies in the MHz range, as in this study, hydrophones are fragile to damage from cavitation induced by the acoustic field. 
Finally, due to the spatial averaging effect, the pressure field obtained in experiments is averaged over a finite local area.
The fluctuating pressure field, which is smaller than the averaging area, cannot be measured.

The spatial averaging effect of hydrophones has been extensively studied.
A well-established equation (Eq. \ref{eq:hyd_effective_dia}) defines the effective diameter of the hydrophone, $d_h$, for conditions where the spatial averaging effect can be minimized \cite{harris1985discussion}.
This equation assumes that the hydrophone is aligned along the central axis of the focused acoustic field \cite{harris1985discussion}.
Note that in cases where both the ultrasound and the hydrophone are arranged in a planar, parallel configuration, the spatial averaging effect is generally negligible.
When ultrasound waves propagate from a circular transducer, they follow two distinct paths, $l_1$ and $l_2$, as shown in Fig. \ref{fig:hydrophone} (b).
The spatial averaging effect is caused by the path differences $l_2$ - $l_1$, which result in the detection of ultrasound waves of different phases at the piezoelectric element surface.
Thus, reducing the hydrophone’s diameter decreases the difference $l_2$ - $l_1$, reducing the spatial averaging effect.
Harris \cite{harris1985discussion} proposed a criterion for $d_h$ based on the assumption that the spatial
averaging effect is negligible when $l_2$ - $l_1$ < $\lambda / 4$, where $\lambda$, where $\lambda$ is the wavelength of the target ultrasound.
This criterion is expressed as:
\begin{equation}
    d_h< 0.5 \frac{\lambda l_h}{d_s},
    \label{eq:hyd_effective_dia}
\end{equation}
where $l_h$ is the distance from the center of the transducer to the hydrophone tip, and $d_s$ denotes the diameter of the transducer (see Fig.\ref{fig:hydrophone}(b)).








\section{Experiments}\label{sec:exp}
\subsection{Experimental setup}\label{sec:exp-setup}
\begin{figure}[ht]
	\centering
	\includegraphics[width=1.0\columnwidth]{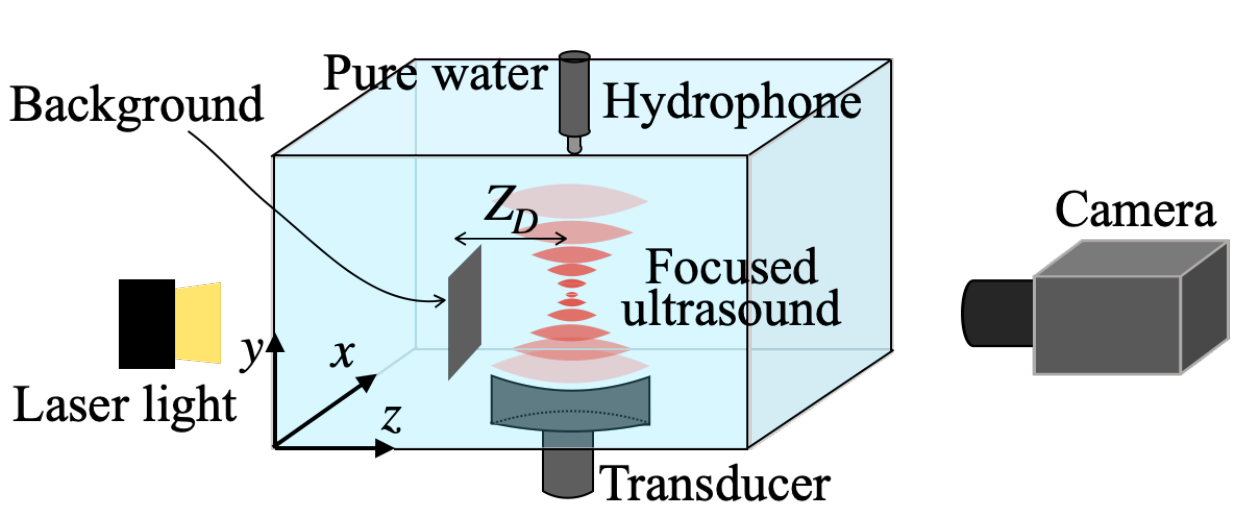}
	\caption{A schematic of the experimental setup. 
    A camera, a transducer, a background, and a light source aligned along a line parallel to the $z$-axis.
    The transducer was positioned at the center of the tank’s bottom, which was filled with pure water.
    This transducer generated a focused acoustic field (frequency: 4.55 MHz) in water.
    The distance between the background and the focal region of the acoustic field, $Z_D$, was set to 21 mm.
    A hydrophone was employed to measure the pressure simultaneously with the BOS measurements.
    The hydrophone tip was positioned in the focal region of the focused ultrasound, where the pressure amplitude is at its maximum, aligned along the central axis.}
	\label{fig:exp-setup}
\end{figure}
The main components of our experimental setup are illustrated in Fig. \ref{fig:exp-setup}.
The experimental setup includes a camera (Canon EOS R5, image size: 8192 $\times$ 5464 pixels), a transducer, a background (20 µm checkerboard pattern), and a light source (SILUX640, Specialized Imaging, wavelength 640 nm, pulse width 10 ns) aligned along a line parallel to the $z$-axis.
A $150 \times 150 \times 150 $ mm$^3$ acrylic tank was filled with pure water.
A transducer with a diameter $d_s$ of 40 mm and a curvature radius of 40 mm (PZT, 4Z40DS40R-Q(C-213), Fuji Ceramics, Shizuoka, Japan) was positioned at the center of the tank's bottom.
The transducer generated a focused acoustic field in water, with a focal distance of approximately 40 mm.
The pressure amplitude of the focal region across the $xz$ plane followed a two-dimensional Gaussian distribution.
A multifunction generator (NF, WF1974) was used to produce a 4.55 MHz sine wave signal with a 0.25 V amplitude, which was amplified using an amplifier (Mini-Circuits, LZY-22+) before being applied to the transducer.

The camera was focused on the background.
The focal region of the ultrasound field was centered within the measurement area (5 × 8 $\rm{mm}^2$).
The distance between the background and the focal region of the acoustic field, $Z_D$, was set to 21 mm to prevent background interference with the acoustic field, ensuring that $Z_D$ exceeded the transducer's radius of 20 mm.
$Z_D$ is a critical parameter that influences the sensitivity of BOS measurements (Eqs. \ref{eq:bos_u} and \ref{eq:bos_v}).
The method for determining the optimal $Z_D$ under experimental restrictions will be discussed in Section \ref{sec:result_disp-p}.
High-speed imaging was achieved by synchronizing the multifunction generator, camera, and light source using a delay generator (Model 575, BNC).
The camera shutter opened simultaneously with the transducer's activation.
Herein, the time of transducer activation is defined as $t = 0$. 
After the transducer was activated, the light source emitted a 10 ns pulse, and the camera shutter then closed.
The camera captured images between $t =$ 32,000 and 32,660 ns from the transducer activation to light activation in increments of approximately 15 ns.
Approximately 100 image sets were captured for each light activation time, each consisting of a reference image and a distorted image.
Applying FCD to the reference and distorted images at a specific time resulted in the displacement fields shown in Fig. \ref{fig:fcd}.
A time lag of several nanoseconds from the preset timing sometimes occurred due to variations in the activation time of the light and the transducer.
Five sets were selected from the 100 captured images to minimize these effects.
This selection method is detailed in Section \ref{sec:exp-analysis}.

A hydrophone was employed to measure the pressure simultaneously with the BOS measurements.
Two hydrophones with diameters of 1.2 mm (M\"uller--Platte Needle Probe, Mueller Instruments) and 2.2 mm (HNR-0500, ONDA Corporation) were utilized separately.
The hydrophone tip was positioned in the focal region of the focused ultrasound, where the pressure amplitude is at its maximum, aligned along the central axis.
The distance $l_h$ between the hydrophone tip and the transducer was placed near the ultrasound focal region.
Based on Eq. \ref{eq:hyd_effective_dia} and the ultrasound wavelength $\lambda$ of 329 µm in this experiment, a hydrophone with a diameter $d_h$ of less than 165 µm should be free from spatial averaging effects.
Since both hydrophones had a piezoelectric element size of 0.5 mm, which is much larger than 165 µm, the spatial averaging effect could not be avoided for either hydrophone type.

\subsection{Data analysis procedure}\label{sec:exp-analysis}

\begin{figure*}[ht]
	\centering
	\includegraphics[width=1.0\textwidth]{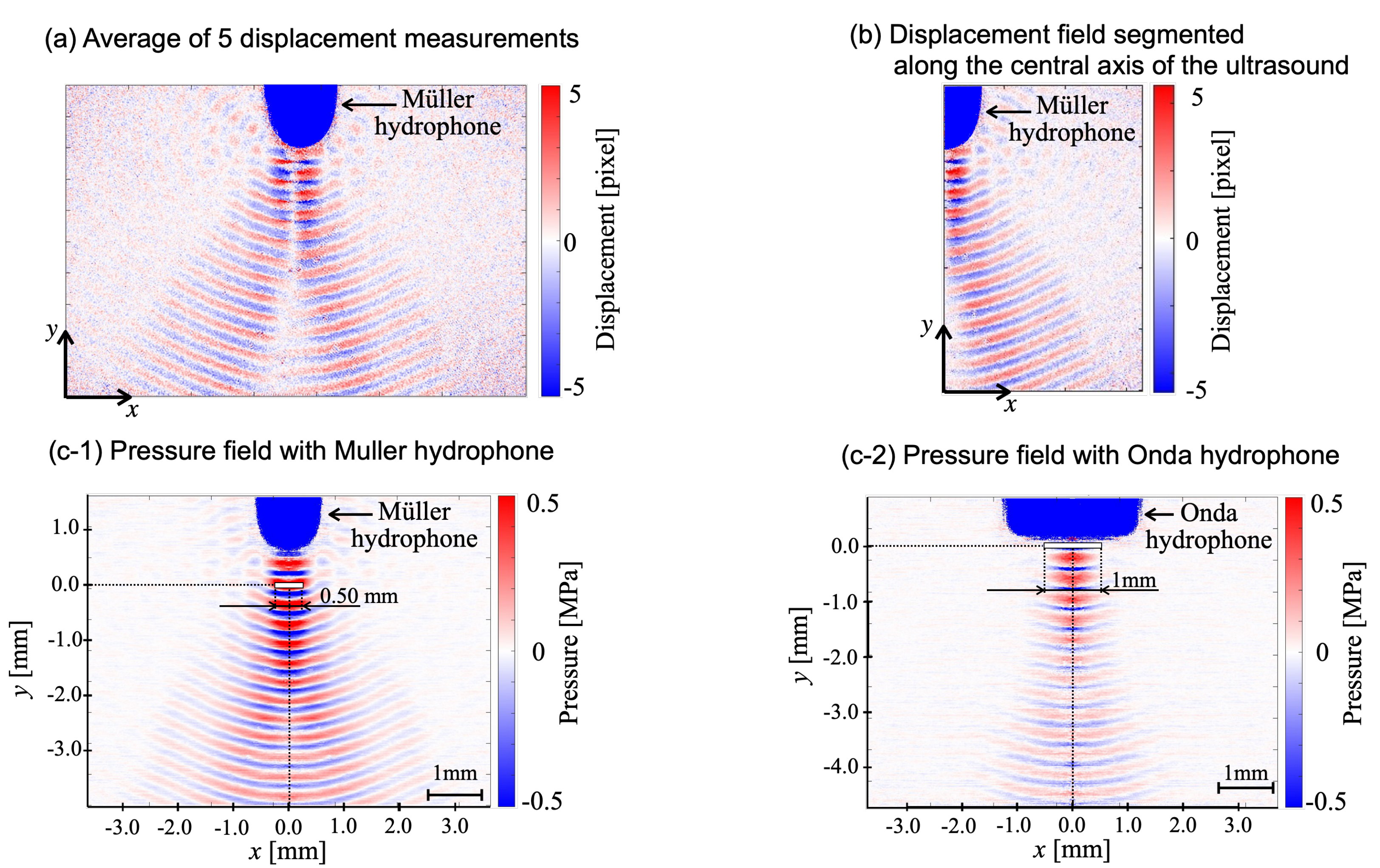}
	\caption{
 The data analysis procedure for calculating the pressure field from displacement measurements.
 (a): The averaged $x$-displacement field measured by FCD.  
 (b):  After dividing along the symmetry axis, the $x$-displacement (a) was averaged from both sides. Note that, one side was flipped along the symmetry axis and its sign was reversed before averaging.
 (c-1) and (c-2): The focused acoustic pressure fields calculated from the displacement field using VT-BOS, with the M\"uller hydrophone and Onda hydrophone inserted, respectively.
 To compare the results of the hydrophone and BOS, the values averaged over the white region are considered the measurements obtained by BOS. 
 The width of the white region is determined to be half of the hydrophone diameter (0.5 mm in (c-1) and 1 mm in (c-2)), while the height is set to 3.3 µm, corresponding to the product of the hydrophone’s temporal resolution and the sound speed.}
	\label{fig:exp-analysis}
\end{figure*}

This section details the data analysis procedure for calculating the pressure field from displacement measurements.
The process is as follows: (1) Five displacement fields are selected from 100 data sets under the same experimental conditions.
(2) The axis of symmetry is defined, which is essential for the VT assumption (Sec. \ref{sec:vt}).
(3) After VT, the boundary conditions are set when integrating the density gradient $\partial \rho/ \partial r$ obtained from the reconstruction.
(4) The results obtained from the hydrophone and those from VT-BOS are compared.

(1) In this experiment, a time lag of a few nanoseconds may have occurred between the 100 data sets at the same light activation time (see Section \ref{sec:exp-setup}).
To minimize the impact of this lag, the 100 displacement fields were averaged. 
Then, each displacement field was compared to the averaged one. 
Based on these comparisons, five displacement fields with ultrasound phase differences of less than 1 µm from the averaged field were selected.
Figure \ref{fig:exp-analysis}(a) illustrates the average of the five displacement fields.
The exact $t$ of the averaged displacement field is estimated from the travel distance of the ultrasound divided by the sound speed (1497.4 m/s).

(2) The averaged displacement field was used in the three-dimensional reconstruction method, VT.
To achieve accurate calculation using VT, the position of the symmetry axis of the acoustic field must be determined beforehand.
In symmetrical measurement targets, the displacement fields along the symmetry axis are either inverted or symmetric between both sides.
We used this property to define a symmetry axis.
First, the $x$-displacement field on one side of a tentative axis was flipped and then added to the corresponding displacement field on the opposite side ($S_1$). 
Similarly, the $y$-displacement field on one side was inverted along the tentative axis and subtracted from the displacement field on the other side ($S_2$).
By scanning the position of the tentative axis, the axis with the smallest combined value of $S_1$ and $S_2$ was identified as the symmetry axis.
As shown in Fig. \ref{fig:exp-analysis}(b), the averaged $x$-displacement from both sides, after dividing along the symmetry axis, was used in the VT.
Before averaging, the displacement field on one side was flipped and its signs were reversed.
Finally, VT (Eq. \ref{Eq:VT-bos}) was applied to the $x$-displacement field to calculate the density gradient $\partial \rho / \partial r$ of the focused acoustic field.

(3) The density field was obtained by integrating the density gradient field, $\partial \rho / \partial r$, over space.
The $r$-axis in polar coordinates corresponded to the $x$-axis in Cartesian coordinates, given that $\theta$ = 0.
The density of water, $\rho_0$, was used as a boundary condition because the change in the density gradient caused by the focused ultrasound was assumed to be negligible at locations far from the $y$-axis. 
In this experiment, the furthest $y$-axis location from the symmetry axis was at $x$ = 3.6 mm (though ideally, $x = \infty$) where the condition $\rho = \rho_0$ was assumed in Eq. \ref{Eq:bos_rhog}.
The density gradient field $\partial \rho / \partial r$ was then integrated along the $x$-axis from 3.6 mm toward 0 mm to compute the density field.
Subsequently, Tait's equation (Eq.\ref{eq:bos_tait}) was applied to this density field to obtain the pressure field in water, as depicted in Fig. \ref{fig:exp-analysis}(c-1) and (c-2).

(4) The pressure field measured by VT-BOS exhibited finer spatial resolution than the field measured by hydrophones.
As noted in Sec. \ref{sec:exp-setup}, hydrophone measurements are significantly affected by the spatial averaging effect due to the relatively large diameter and sensor size of the hydrophones, which exceed the effective diameter $d_h$ required to minimize such effects.
To facilitate a direct comparison between VT-BOS and hydrophone measurements, the pressure field obtained by VT-BOS was averaged over an area shown by the white region in Fig. \ref{fig:exp-analysis}(c-1) and (c-2).
The vertical width of this region was determined by multiplying the hydrophone's time resolution of 1 ns by the speed of sound in water (1497.4 m/s).
The horizontal width was set according to the hydrophone radii: 0.5 mm for the M\"uller hydrophone and 1.0 mm for the Onda hydrophone.

\section{Results and discussion}\label{sec:result}

In this section, we discuss the spatiotemporal measurement results for the focused acoustic pressure field ob- tained using VT-BOS.
Section \ref{sec:result_p-field} presents the acoustic pressure fields measured in static water, both with and without hydrophones.
Section \ref{sec:result_comparison} compares the results obtained by VT-BOS with those from other methods.
Section \ref{sec:result_t-p} specifically compares the temporal evolution of the local pressure as measured by VT-BOS and hydrophones.
The discussion on the spatial averaging effect, which affects the pressure amplitude measured by hydrophones, highlights the high-resolution measurement capabilities of VT-BOS.
Section \ref{sec:result_cost} examines differences in the computational time and spatial resolution of VT-BOS and other measurement techniques for spatiotemporal acoustic field measurements.
Through these comparisons, we demonstrate the relative effectiveness of VT-BOS in focused acoustic measurements.
Finally, Section 4.3 discusses the limitations of the proposed technique under the specific experimental conditions used in this study.

\subsection{Spatiotemporal pressure field}\label{sec:result_p-field}

\begin{figure*}[ht]
	\centering
    \includegraphics[width=1.0\textwidth]{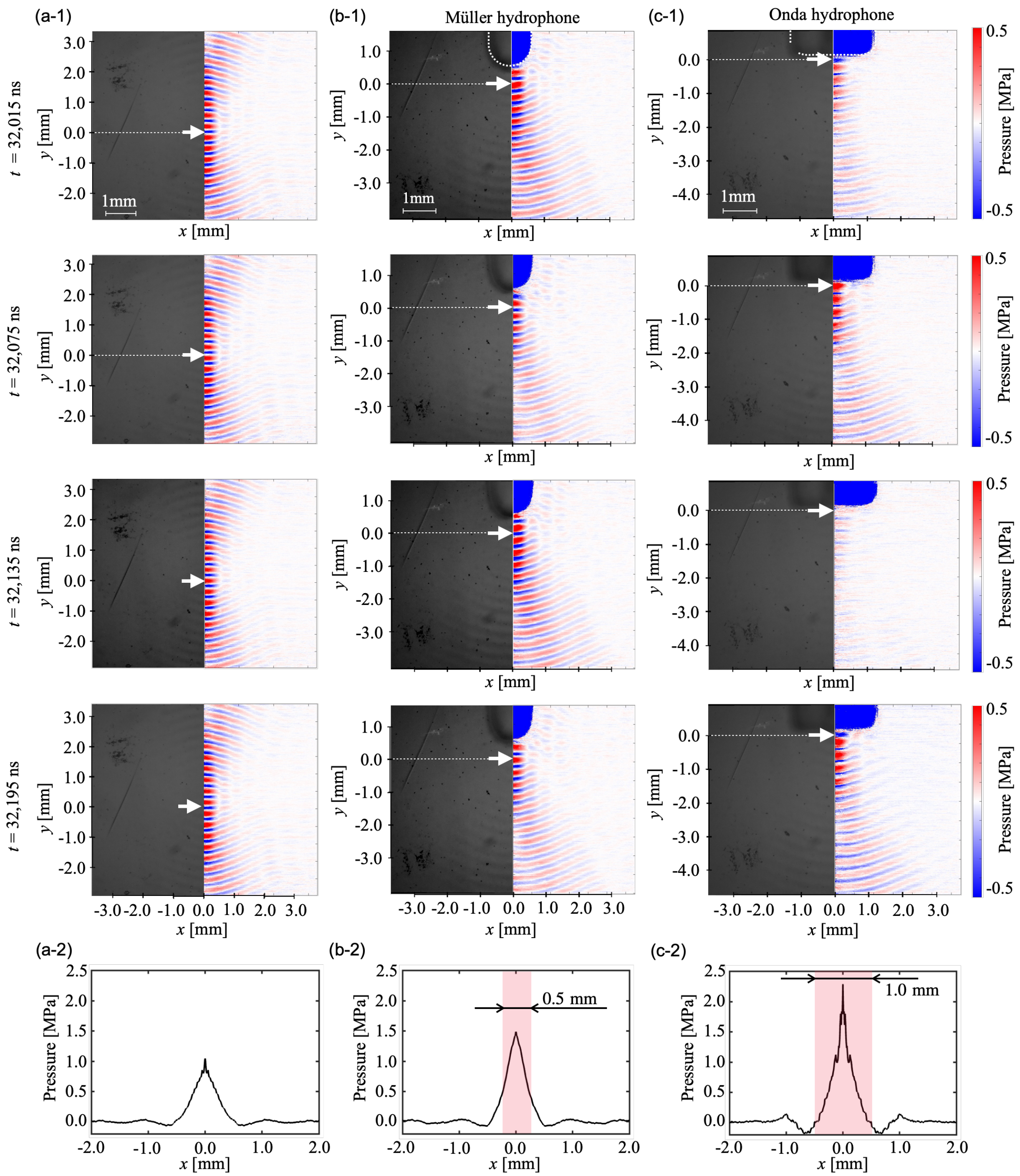}
	\caption{
    The time evolution of the distorted images and the acoustic pressure fields measured by VT-BOS, and the maximum pressure amplitude in the focal region are presented. 
    (a-1), (b-1), and (c-1) The background images (left panel) and the pressure fields measured by the VT-BOS technique (right panel) at t = 32,015, 32,075, 32,135, and 32,195 ns.
    The acoustic field depicted in (a-1) includes no hydrophones (scatterers).
    The acoustic fields shown in (b-1) and (c-1) include M\"uller and Onda hydrophones, respectively.
    The white arrow indicates the focal point of the ultrasound at (0,0).
    (a-2), (b-2), and (c-2) show the the acoustic pressure along the $x$-axis at $y =$ 0.0 mm, as shown by the white dotted line in (a-1), (b-1), and (c-1).
    The pressure at time $t$  when the maximum amplitude is obtained is shown in (a-2), (b-2), and (c-2).
    The pink regions in (b-2) and (c-2) represent the horizontal width ($x$-direction) of the averaged area based on the hydrophone diameter.
    }
	\label{fig:p-field}
\end{figure*}

The time evolution of the distorted images and the acoustic pressure fields measured by VT-BOS are presented respectively in the left and right panels of Fig. \ref{fig:p-field} (a-1), (b-1), and (c-1).
We first focus on the pressure field of the temporal evolution and then discuss the maximum pressure amplitude in the focal region.

The pressure field without a hydrophone, as measured by VT-BOS, is illustrated in Fig. \ref{fig:p-field} (a-1).
A significant pressure amplitude is identified in an hourglass-shaped area near $x$ = 0.0.
The pressure field alternates between positive and negative values along the $y$-axis, with the pressure amplitude increasing toward the focal region.
The white arrow in the figure indicates the position (0, 0) where the maximum pressure amplitude occurs (i.e., the focal position of the ultrasound).
The pressure amplitude at this position varies with time.
Comparing the pressure amplitude at (0,0) at $t$ = 32,015 ns and $t$ = 32,135 ns, there is approximately a half-wavelength phase shift.
From the BOS results, the period is estimated as $(32,135 - 32,015) \times 2 = 240$ ns.
Assuming a sound velocity of 1500 m/s, the period corresponding to a frequency of 4.55 MHz is approximately 220 ns.
The BOS measurements confirm that ultrasound propagates through water at the expected sound velocity. 
A detailed comparison, including period analysis, will be provided in Section \ref{sec:result_t-p}.
In the measurement region where 
$x$ \textgreater \ 1 mm, non-zero pressure fluctuations are clearly observed, and at $x$ \textgreater \ 3 mm, these fluctuations can be considered negligible, approximating atmospheric pressure.

The pressure field obtained from BOS, which is disturbed by the M\"uller hydrophone is shown in Fig. \ref{fig:p-field} (b-1).
The pressure field exhibits alternating positive and negative pressures along the $y$-axis, with the pressure amplitude increasing toward the focal region around (0,0).
For the pressure amplitude at (0,0), the phase of the pressure wave varies with time.
The amplitude of the pressure field along the $y$-axis at $x$ = 0.0 also changes its pattern with time, exhibiting a different trend to the results shown in Fig. \ref{fig:p-field} (a-1).
Notably, the M\"uller hydrophone which has a diameter of 1.0 mm and a round tip, affects the pressure field in its vicinity.
The traveling wave interacting with the M\"uller hydrophone not only reflects in the direction of propagation but is also scattered in multiple other directions.

The pressure field disturbed by the Onda hydrophone is shown in Fig. \ref{fig:p-field} (c-1).
There are negligible pressure fluctuations near the Onda hydrophone, which differs from the observation near the M\"uller hydrophone in Fig. \ref{fig:p-field} (b-2).
The Onda hydrophone has a diameter of approximately 2.0 mm and a flat tip.
The traveling wave at the flat tip of the Onda hydrophone predominantly reflects in the direction of propagation.
The pressure amplitude at (0,0), varies with time. 
Notably, at times $t$ = 32,015 -- 32,195 ns, the pressure field exhibits a very small phase shift.
This indicates that a standing wave is created by the interference between the incident wave and the reflected wave from the Onda hydrophone in the pressure field.
\clearpage
\clearpage

For a more quantitative comparison, the pressure distri- bution along the $x$-axis at $y$ = 0.0 mm is shown in Fig. \ref{fig:p-field}(a-2), (b-2), and (c-2).
These figures show the pressure fields for cases without hydrophones, with a M\"uller hydrophone, and with an Onda hydrophone, respectively.
The amplitude decreases in each of these figures as the distance from $x = 0.0$ increases.
The width of the focused pressure field along the $x$-axis is approximately 1.0 mm in all three cases.
The peak amplitudes of 1.05 MPa, 1.49 MPa, and 2.29 MPa are observed at $x$ = 0.0 for Fig. \ref{fig:p-field} (a-2), (b-2), and (c-2), respectively.
Notably, the peak amplitude in Fig. 8 (c-2) is approximately twice that in Fig. \ref{fig:p-field} (a-2).
These results suggest that a standing wave was generated due to the interference of the incident and reflected waves by the Onda hydrophone, as indicated in Fig. \ref{fig:p-field} (c-1).
In contrast, a traveling wave was measured in Fig. \ref{fig:p-field} (a-1).
In the acoustic field of Fig. \ref{fig:p-field} (b-1), the wave reflected by the M\"uller hydrophone was scattered in all directions, in contrast to the reflection pattern observed in Fig. \ref{fig:p-field} (c-1).
This difference in reflection patterns explains why the peak amplitude of Fig. \ref{fig:p-field} (b-2) is lower than that of Fig. \ref{fig:p-field} (c-2) but higher than that of Fig. \ref{fig:p-field} (a-2).
Note that VT-BOS may increase the measurement error around the axis of symmetry (the $y$-axis) as described by Ichihara $\it{et \ al}$. \cite{ichihara2022background}, which may lead to the peak amplitude in Fig. \ref{fig:p-field} (c-2) (2.29 MPa) being slightly larger than twice the value in Fig. \ref{fig:p-field} (a-2) (1.05 MPa).

\subsection{Comparison}\label{sec:result_comparison}
\subsubsection{Temporal pressure}\label{sec:result_t-p}
\begin{figure}[ht]
	\centering
	\includegraphics[width=1.0\columnwidth]{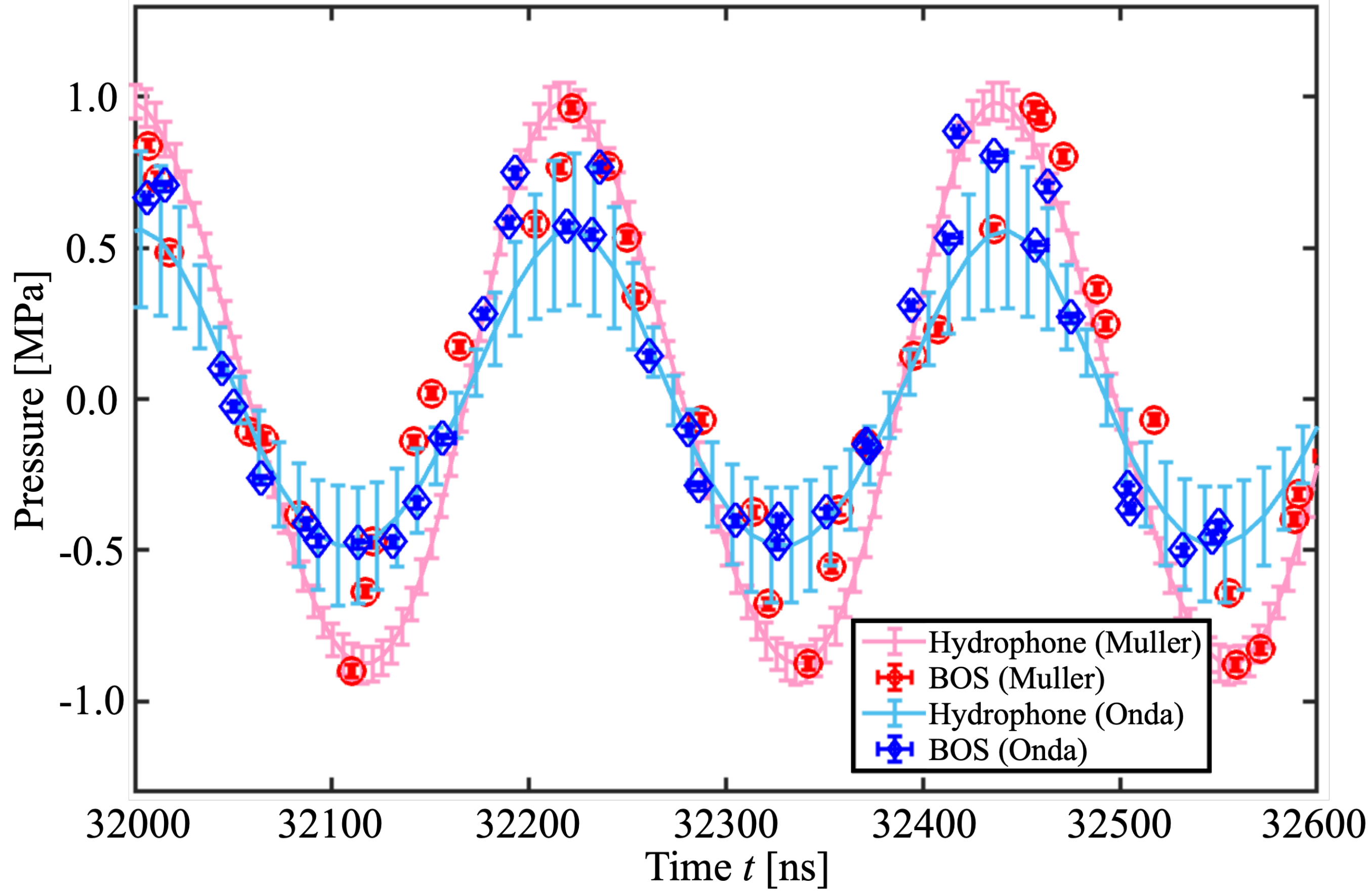}
	\caption{
    Pressure values of the time-evolving acoustic field at a specific position.
    The solid lines represent the values measured by the hydrophones, while the markers indicate the results obtained by BOS.
    Red represents the acoustic field from the M\"uller hydrophone, and blue represents the acoustic field from the Onda hydrophone. The red markers represent the averaged pressure field over the region $x$ = 0 $\pm$ 0.25 mm and $y$ = 3.8 mm $\pm$ 11.2 µm.
    The blue markers represent the averaged pressure field over the region $x$ = 0 $\pm$ 0.50 mm and $y$ = 3.8 mm $\pm$ 11.2 µm.}
	\label{fig:t-p}
\end{figure}

In this subsection, we first compare the measured frequencies of pressure fluctuations to assess the accuracy of each measurement method.
Following this, we directly compare the temporal local pressure measurements from VT-BOS and the hydrophones, as shown in Fig. \ref{fig:t-p}.
Since the pressure field measured by BOS has a much higher spatial resolution than the M\"uller or Onda hydrophones, the local pressure values of the BOS results are defined as the average value over an area proportional to the hydrophone radius, as shown by the white area in Fig. \ref{fig:exp-analysis} (c-1) and (c- 2), respectively.
In Fig. \ref{fig:t-p}, the red circles and blue diamonds represent the pressure value fields from the M\"uller and Onda hydrophones, hereafter referred to as BOS(M\"uller) and BOS(Onda), respectively.
The solid red and blue lines depict the measurements from the M\"uller and Onda hydrophones, referred to as M\"uller and Onda, respectively.

The calculated frequency was 4.50 MHz for both hydrophones and 4.44 MHz for the BOS measurements. The transducer generated focused ultrasound at 4.55 MHz.
Considering the time resolution of 15 ns for BOS and 1 ns for the hydrophones, the measured frequencies are in agreement, indicating that the ultrasound frequency measurements are reasonable.

The pressure amplitude measured by the Onda hydrophone (the blue solid line) is substantially lower than that measured by the M\"uller hydrophone (the red solid line), although the input ultrasounds are the same.
The outer diameters of the M\"uller and Onda hydrophones are approximately 1.0 mm and 2.0 mm, respectively, while the sensor size of both hydrophones is 0.5 mm.
It is important to note that the effective diameter is less than 165 µm according to Eq. \ref{eq:hyd_effective_dia} (Section \ref{sec:hydrophone}), which is significantly smaller than both the sensor size and the hydrophone diameters. 
Thus, the spatial averaging effect may be responsible for the difference in pressure amplitudes measured by the M\"uller and Onda hydrophones.

\begin{table*}[!ht]
\caption{
A summary comparing the proposed VT-BOS technique with conventional BOS, schlieren, phase contrast imaging, and hydrophone methods in terms of area, spatial resolution, computational time, required measurement time, and maximum measurable amplitude for spatiotemporal 3D acoustic field measurements.
Hyphens in the computational and measurement time table indicate that, to our knowledge, conventional BOS and schlieren techniques no examples of spatiotemporal measurements. Additionally, the hyphen for hydrophone computational time signifies that it is unnecessary.}
\label{tb:cost}
\resizebox{\textwidth}{!}{
\footnotesize
\begin{tabular}{rcccc}
\toprule
  &\begin{tabular}{c} Spatial resolution \\ (image size)\end{tabular}    &Computational time & Measurement time&Amplitude\\ 
\midrule
VT-BOS\cite{ichihara2022background}& 
\begin{tabular}{c} 1 µm \\ (8192 $\times$ 5464 pixels) \end{tabular}
& 14 s & $\le$ 6 mins & 0.4 - 6.5 MPa\\
Conventional BOS\cite{luo2020rapid, koponen2022nonlinear} &
\begin{tabular}{c} 425 µm \\ (4000 $\times$ 6000 pixels) \end{tabular}
& - & - & 9 MPa\\
Schlieren \cite{neumann2006schlieren,colom2023rapid} & 
\begin{tabular}{c} 26 µm \\ (1390 $\times$ 1024 pixels) \end{tabular}
& - & - & 1 MPa\\
Phase contrast imaging\cite{nakamura2018quantitative}  &
\begin{tabular}{c} 47 µm \\ (1400$\times$ 1400 pixels) \end{tabular}
& 12 hours & 30 mins& $\sim$ 0.2 MPa\\
Hydrophone\cite{nakamura2018quantitative,kothapalli2017acoustic} & 100 µm & - & 1 month &  1 GPa\\
\bottomrule
\end{tabular}
}
\end{table*}


The pressure values measured by VT-BOS, both BOS(M\"uller) and BOS (Onda), generally align with the measurements from the M\"uller and Onda hydrophones. 
Remarkably, the BOS(M\"uller) and BOS(Onda) values around the pressure nodes closely match the corresponding hydrophone measurements within the errors.
This indicates that the M\"uller and Onda hydrophones measure averaged pressures over areas of 0.5 mm and 1.0 mm in width, as shown by the white areas in Fig.  \ref{fig:exp-analysis} (c).
The BOS results indicate that the spatial averaging effect is closely related to the hydrophone’s outer radius under these conditions.

It is noteworthy that the amplitude along the horizontal axis toward the focal direction in BOS(Onda) is larger than that in BOS(M\"uller), as shown in Fig. \ref{fig:p-field} (b-2) and (c-2).
This trend is the opposite of what is observed in Fig. \ref{fig:t-p}.
This is attributed to the difference in the averaging areas of BOS(M\"uller) and BOS(Onda).

The above discussion on the spatial averaging effect, which affects the pressure amplitude measured by hydrophones, highlights the high-resolution measurement capabilities of VT-BOS.
The high-resolution VT-BOS technique has significant potential for measuring ultrasound on fine scales, e.g., in microfluidics, and for estimating the spatial averaging effect in hydrophones as a calibration tool for hydrophone measurements.


\subsubsection{Performance comparisons}\label{sec:result_cost}

In this section, we compare the proposed VT-BOS with the conventional BOS, schlieren, phase contrast imaging, and hydrophone techniques.
We focus on the measurement area, spatial resolution, computational time, required measurement time, and maximum measurable amplitude for spatiotemporal 3D acoustic field measurements.
These comparisons are summarized in Table \ref{tb:cost}.
The results for each method are based on previous studies on ultrasound measurements cited in the table.
To the best of our knowledge, there are no examples of spatiotemporal measurements of ultrasound using conventional BOS and schlieren techniques; therefore, these methods have been excluded from the computational time and measurement time categories.

First, we discuss the spatial resolution of each technique.
In non-contact measurement techniques, the spatial resolution depends on the camera magnification and the pixel size of the image sensor.
In our experiment, a DSLR camera with a high-pixel imaging sensor was used, enabling higher spatial resolution measurements than were achieved in previous studies.
However, for hydrophones, spatial resolution is determined by the sensor size, with the most recent advances achieving a minimum spatial resolution of 40 µm with PVDF hydrophones \cite{harris2022hydrophone}.

Second, we compare VT-BOS, phase contrast imaging, and hydrophone techniques in terms of computational and measurement time for 3D acoustic field measurements. 
Hydrophone measurements require extended time for 3D pressure field mapping due to the need for scanning.
In contrast, VT-BOS and phase contrast imaging offer significantly shorter computational and measurement times as they rely on image-based measurement techniques.
For example, VT-BOS directly calculates the reconstructed density gradient from the vector field (Eq.\ \ref{Eq:VT-bos}), eliminating the need to solve the Poisson equation and substantially reducing the computational time.
A more detailed discussion of this computational efficiency can be found in Ichihara $\it{et \ al.}$ (2022) \cite{ichihara2022background}.

Finally, the measurable maximum amplitudes differ significantly among the five techniques.
Hydrophones can measure the highest maximum pressure amplitude among the considered methods, reaching 1 GPa.
VT-BOS was able to measure pressures ranging from 0.4 to 6.5 MPa in this experiment (as discussed further in Section \ref{sec:result_disp-p}), which is a similar range to that of the schlieren and conventional BOS techniques.
In contrast, phase contrast imaging can measure pressures lower than 0.2 MPa by utilizing light phase differences.
VT-BOS is suitable for measuring focused acoustic fields at frequencies of several MHz and pressures of several MPa, unlike the other methods.

\subsection{Measurement limitations and prospects of VT-BOS}\label{sec:result_disp-p}

\begin{figure}[ht]
	\centering
	\includegraphics[width=1.0\columnwidth]{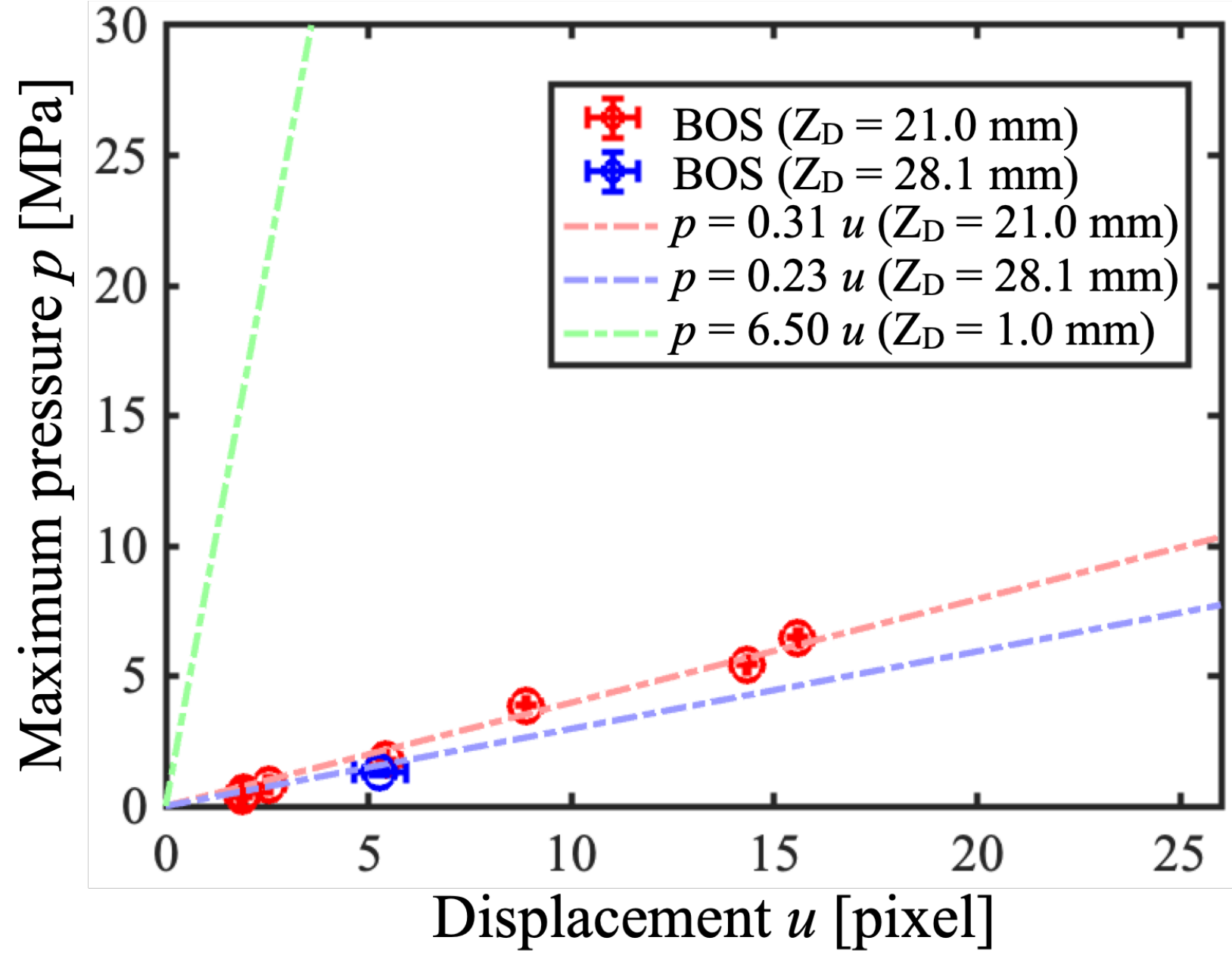}
	\caption{
    The relationship between the maximum pressure amplitude $p$ in the acoustic pressure field and maximum $x$-displacement $u$ measured by BOS.
    The red markers indicate the results in the different maximum pressure amplitude increments at $Z_D$ =21 mm.
    The blue marker indicates the experimental result for the maximum displacement and maximum amplitude at $Z_D$ =28.1 mm and $V$ = 0.25 V.
    The red dashed line represents the linear approximation obtained by the least-squares method from the red markers' results.
    The blue and green dashed lines represent theoretical estimation at $Z_D$ =28.1 mm and $Z_D$ =5.0 mm, respectively.}
	\label{fig:disp-p}
\end{figure}

As a final subsection, we discuss the limitations and prospects of VT-BOS when applied to measure underwater ultrasound.
The measurable maximum pressure amplitude in BOS is primarily determined by the displacement calculated using FCD under specific experimental conditions, as suggested by Eq. \ref{Eq:VT-bos}.
Figure \ref{fig:disp-p} illustrates the relationship between the maximum pressure at the focal region and the corresponding maximum displacement.
The red markers represent the BOS measurement results at $Z_D$ = 21 mm, while red dashed line indicates the linear relationship between displacement and pressure, as calculated using the least squares method.
The red markers show that VT-BOS successfully quantified the maximum pressure amplitude from 0.4 to 6.5 MPa.
In this experiment, the checkerboard pattern in the background had a square size of 17.5 pixels, with a wavenumber $k_c = \pi / (17.5\sqrt{2}) = 0.13$ pixel$^{-1}$.
Based on this, the maximum displacement $u_{\text{max}}$ in this pattern size is 24.8 pixels, as determined by the FCD criterion (Eq. \ref{Eq:fcd-criteria1}).
The red dashed line ($p = 0.40 \ u$) represents the linear correlation between the displacement and pressure, as calculated from the red markers.
The displacement is proportional to the density gradient (as per BOS theory in Eq. \ref{Eq:VT-bos}), and the density-pressure relationship follows the Tait equation, thus explaining the linear relationship between displacement and pressure.

In BOS, the displacement $u$ depends on the parameter $Z_D$, as described by Eq. \ref{eq:bos_u}.
By adjusting $Z_D$, the measurement sensitivity of BOS can be optimized.
Increasing $Z_D$ enhances the ability of the method to measure lower pressures (below 1 MPa), while decreasing $Z_D$ is beneficial for measuring higher pressures (above 10 MPa).
The blue marker in the figure represents an experimental result at $Z_D = 28.1$ mm, showing a displacement of approximately 5 pixels for a pressure of 1.2 MPa.
In contrast, the red markers at $Z_D = 21$ mm show a displacement of around 4 pixels for a pressure of 1 MPa.
The ratio of displacements between these two conditions (4 pixels/5 pixels = 0.80) closely matches the ratio of their $Z_D$ values (21/28.1 = 0.75), indicating that the displacement scales proportionally with $Z_D$ under similar measurement conditions.

The blue ($p = 0.30 u$) and green ($p = 8.40 u$) dashed lines in the figure represent the displacement-pressure relationships at $Z_D = 28.1$ mm and $Z_D = 1.0$ mm, respectively.
The slopes $a_{\text{blue}}$ and $a_{\text{green}}$ are derived from the red dashed line ($p = 0.40 u$), with $a_{\text{blue}} = 0.40 \times (21/28.1) = 0.30$, and $a_{\text{green}} = 0.40 \times (21/1.0) = 8.40$.
These relationships were established by keeping all experimental conditions constant, except for $Z_D$.
The maximum and minimum $Z_D$ values in these experiments were set to 28.1 mm and 1.0 mm, respectively.
This choice was made because, for $Z_D > 28.1$ mm, the width of the light ray at the target section becomes significant, leading to spatial averaging of the density gradient, which can affect measurement accuracy \cite{gojani2013measurement}.
For $Z_D < 1.0$ mm, the background would fully overlap with the focused acoustic field, disturbing the flow field.

Sub-pixel displacement values were evaluated using Gaussian interpolation in FCD, which has known limitations and may not be entirely reliable \cite{wildeman2018real}.
Therefore, in this experimental setup, we can confidently rely on displacement values ranging from 1 to 24 pixels.
Considering these limitations and the FCD measurement range, the BOS technique can measure pressures between 0.3 MPa (corresponding to 1 pixel) and 201.6 MPa (corresponding to 24 pixels) if $Z_D$ is varied from 1.0 mm to 28.1 mm. For measuring low-pressure fluctuations of several pascals, phase contrast imaging may offer a more suitable alternative.

A key limitation of VT-BOS is its assumption that the measurement target is axisymmetric and repeatable.
For instance, acoustic fields generated by horn transducers at frequencies of several hundred kHz, often used in sonochemistry and sonoluminescence \cite{merouani2024review}, may produce non-axisymmetric fields and highly unsteady phenomena such as cavitation.
Several strategies can be considered to overcome these challenges.
Increasing the number of camera viewing angles and employing the algebraic reconstruction technique (ART) may allow for the measurement of non-axisymmetric targets \cite{amjad2020assessment,amjad2023three}.
Additionally, using high-speed cameras enables BOS to capture unsteady phenomena.

In cases where the measurement target contains small scatterers, such as bubbles or solid particles, light scattering can negatively affect BOS performance.
To mitigate this, BOS can be applied to acoustic fields with reduced bubble formation, can focus on the pressure field just before cavitation, or can be used primarily as a visualization tool rather than for quantitative analysis.

Finally, the application of BOS to acoustic field measurements in opaque objects, such as the human body, poses a significant challenge.
However, combining BOS with X-ray technology, which can penetrate non-transparent objects, may lead to the development of a technique capable of measuring acoustic fields within such materials.

\section{Conclusion}\label{sec:conclusion}
This study has demonstrated that BOS, enhanced with FCD and vector tomography (VT), can successfully quantify the spatiotemporal pressure field of a 4.554 MHz, multi-megapascal focused acoustic wave with high spatial resolution (1 µm) during the initial 0.66 µs of irradiation.
A comparison of the acoustic field measurements from VT-BOS and hydrophones revealed that VT-BOS could effectively capture the pressure field ranging from 0.4 to 6.5 MPa, including traveling, reflected, and standing waves generated by the interaction with hydrophones.
When the focal width of the ultrasound was smaller than the hydrophone diameter, VT-BOS demonstrated the ability to estimate the spatial averaging effects of hydrophone measurements.
This suggests that VT-BOS has the potential to serve as a calibration tool for hydrophone-based measurements.
Furthermore, VT-BOS enables adjustment of the measurable pressure amplitude by modifying the distance $Z_D$ between the background image and the target.
Under the current experimental setup, this technique measured pressures ranging from 0.3 MPa to 201.6 MPa.

Given its simplicity, VT-BOS is a practical and versatile technique for pressure measurement in ultrasound applications within the MHz and MPa range.
For example, VT-BOS can be used to observe acoustic streaming at frequencies of 1.5 MHz and pressure levels between 1 and 4 MPa, a phenomenon believed to enhance the effectiveness of therapeutic treatments.

\section*{Acknowledgement}
This work was supported by JSPS KAKENHI Grant Numbers JP23KJ0859, JP23H01600, JP24H00289, the HEIWA NAKAJIMA Foundation, International Academic Collaborative Grant, and JST PRESTO Grant Number JPMJPR21O5, Japan. The authors would also like to thank Mr. Ayumu Ishibashi (Master student, Tokyo University of Agriculture and Technology) for technical assistance with the experiments.

\bibliographystyle{ieeetr}
\bibliography{ref}
\end{document}